\documentclass{article}
\usepackage{amsmath}

\usepackage{graphicx}

\begin{document}

\title{Solving a Delay Differential Equation through Fourier Transform}


\author{Kenta Ohira$^{1}$ and Toru Ohira$^{2}$\\
\small \ $^{1}$Future Value Creation Research Center,\\ 
\small Graduate School of Informatics, Nagoya University, Japan\\
\small\ $^{2}$Graduate School of Mathematics, Nagoya University, Japan
}

\maketitle

\begin{abstract}
In this study, we introduce and explore a delay differential equation that lends itself to explicit solutions in the Fourier-transformed space. Through the careful alignment of the initial function, we can construct a highly accurate solution to the equation. These findings open new avenues for understanding delay systems, demonstrating the efficacy of Fourier transform techniques in capturing transient oscillatory dynamics.
\end{abstract}


\section{Introduction}
In various fields such as biology, mathematics, economics, and engineering, understanding the effects of delays has become an important issue \cite{heiden1979,bellman1963,cabrera_1,hayes1950,insperger,kcuhler,longtinmilton1989a,mackeyglass1977,miltonetal2009b,ohirayamane2000,smith2010,stepan1989,stepaninsperger,szydlowski2010}. Even for simple systems, delays in feedback and/or interactions can induce oscillations and complex behaviors.``Delay Differential Equations" (DDEs) are crucial tools in such research. A representative example is the Mackey--Glass equation \cite{mackeyglass1977}, which exhibits various types of dynamics, including chaos. Although our understanding of delay systems and DDEs has gradually improved (e.g.\cite{taylor}), solving DDEs to obtain solutions is considered very difficult.

Typically, the problem of solving delay differential equations is to find a solution $X(t)$ for $ t \in [0, \infty)$, given the initial function $\phi(t)$ for $t \in [-\tau, 0]$. We generally cannot obtain such solutions explicitly. Thus, the main analysis of the DDEs focuses on asymptotic stabilities and behaviors. Also, a recent analytical approach to simple DDEs uses the Lambert W function \cite{corless}, which enables the expression of both formal and approximate solutions of the equation \cite{shinozaki,pusenjak2017,kentaohira2023}.

In this paper, we take an indirect path to obtain the solution of a DDE through the Fourier transform. Specifically, we examine our recently proposed DDE that has a linear time-dependent coefficient \cite{kentaohira2021,kentaohira2024}. With this DDE, the power spectrum peak of the dynamical trajectory reaches its maximum height when the delay is suitably tuned, indicating frequency resonance. Our focus here, however, is to show that we can solve this DDE explicitly in the Fourier-transformed space. We then numerically take the inverse Fourier transform to construct the solution for $ t \in (-\infty, \infty)$. The particular solution for $ t \in [0, \infty)$, then, can be obtained if we view this solution by identifying the initial function $\phi(t)$ with the above semi-numerically constructed function over $ t \in [-\tau, 0)$. This provides a new approach to solve DDEs, and our DDE is a rare example where the solutions are obtained with high accuracy.

\section{Delay Differential Equation}
We proposed and studied the following delay differential equation \cite{kentaohira2021,kentaohira2024}:
\begin{equation}
{dX(t)\over dt} + a t X(t) = b X(t-\tau)
\label{ddr}
\end{equation}
where $a \geq 0$, $b$, $\tau \geq 0$ are real parameters, with $\tau$ interpreted as the delay.

This equation falls into a class of equations called non-autonomous DDEs, which have been studied in a more general context. Here, again, solving non-autonomous DDEs poses mathematical challenges\cite{Busenberg1984,Ming1990,Ford2002,Gyori2017}. Additionally, we can view equation  (\ref{ddr}) as a slight modification of the well-studied Hayes equation, with only the second term replaced by a linear function of time rather than a constant. However, its behavior is quite distinct from that of the Hayes equation.

We have shown that for equation (\ref{ddr}), oscillatory transient dynamics appear and disappear as the delay increases while maintaining asymptotic stability at 
$X=0$. Also, frequency resonance is observed with the power spectrum peak of the dynamical trajectory reaching its maximum height when the delay is suitably tuned.

We list some properties of the special cases of equation (\ref{ddr}):

\begin{itemize}

\item
With the delay $\tau = 0$ and the initial condition $X(t=0) = X_0$, 
the solution to the equation is as follows:
\begin{equation}
X(t) = X_0 e^{- {1\over 2}a t^2 + b t}.
\label{tau0}
\end{equation}
Thus, this solution exhibits a Gaussian shape with  
its peak at $b/a$. 
Furthermore, with $a=0$, we also note that equation  (\ref{ddr}) represents the annihilation operator equation of the quantum simple harmonic oscillator, with $t$ interpreted as a position rather than time (e.g.\cite{sakurai}). It has the Gaussian ground state as its solution.

\item
The case where 
$a=0$ is a special case of the Hayes equation. In this case, the origin 
$X=0$ is asymptotically stable only within the range of 
\begin{equation}
- {\pi/{2 \tau} }< b <0.
\end{equation}
However, with $a > 0$ with a finite initial function, the asymptotic stability of the origin is kept for all delays. For the finite delays, the solution of equation (\ref{ddr}) typically shows oscillatory dynamics. 

\end{itemize}


For the finite delays, the solution of equation (\ref{ddr})  typically shows oscillatory dynamics. 

\subsection{Solution through the Fourier Transform}

Let us now explore finding a solution for equation (1) in the entire range of 
$t \in (-\infty, -\infty)$ without specifying boundary conditions. We reasonably assume that with 
$a>0$, the solution converges to $0$ for $t \rightarrow \pm \infty$ for finite solutions. Fourier transforming equation (\ref{ddr}) yields:
\begin{equation}
i\omega \hat{X}(\omega) + i a {d\hat{X}(\omega)\over d\omega} = - b \hat{X}(\omega) e^{i \omega \tau}
\label{ft1}
\end{equation}
where
\begin{equation}
\hat{X}(\omega) = \int_{-\infty}^{\infty} e^{i \omega t} X(t) dt
\label{ft2}
\end{equation}

Equation (\ref{ft1}) can be easily solved, and the solution is given by:
\begin{equation}
\hat{X}(\omega) = {\cal{C}} Exp[ {- {1\over 2 a} \omega^2 + {b\over \tau a} e^{i \omega \tau}}] 
\label{ftsol}
\end{equation}
where ${\cal{C}}$ is an integration constant. Thus, we can assert the following:
\vspace{1em}

The finite solution of equation (\ref{ddr}) with $a>0$ in the range $t \in (-\infty, -\infty)$ can be explicitly solved, and the solution is given by (\ref{ftsol}) in the Fourier-transformed space. This solution is unique up to the scale factor ${\cal{C}}$.
\vspace{1em}

This leads to the following:
\vspace{1em}

The finite solution of equation  (\ref{ddr}) with $a>0$ in the range $t \in (-\infty, +\infty)$ can be obtained as follows through the inverse Fourier transform:
\begin{eqnarray}
&&{X_f}(t) = {1\over {2 \pi} }\int_{-\infty}^{\infty} e^{- i \omega t} \hat{X}(\omega) d\omega\nonumber\\
&&= { {\cal{C}}\over {2 \pi} }\int_{-\infty}^{\infty} Exp[ {- {1\over 2 a} \omega^2 + {b\over \tau a} e^{i \omega \tau} - i \omega t}] d\omega\nonumber\\
&&= { {\cal{C}}\over {2 \pi} }\int_{-\infty}^{\infty} Exp[ {- {1\over 2 a} \omega^2 + {b\over \tau a} \cos (\omega \tau)}]\cos({b\over \tau a}\sin(\omega \tau) - \omega t )d\omega\nonumber\\
\label{solx2}
\end{eqnarray}
\vspace{1em}
Again, the solution is unique up to the scale factor ${\cal{C}}$. 

We observe that in the special case of $\tau \rightarrow \infty$, the inverse transform simplifies to the scenario of Gaussian integration. Consequently, we can explicitly express the limit as $\tau$ approaches infinity:
\begin{equation}
\lim_{\tau \rightarrow \infty} {X_f}(t) = {\cal{C}} \sqrt{a \over {2 \pi}} Exp[ {- {1\over 2 } a t^2 }] 
\label{ftsolinfty}
\end{equation}

However, for the general case of $\tau$, analytical integration of (\ref{solx2})
is not feasible. Therefore, we resort to numerical integration, and representative results are depicted in Fig. 1.

\begin{figure}[h]
\begin{center}
\includegraphics[height=8cm]{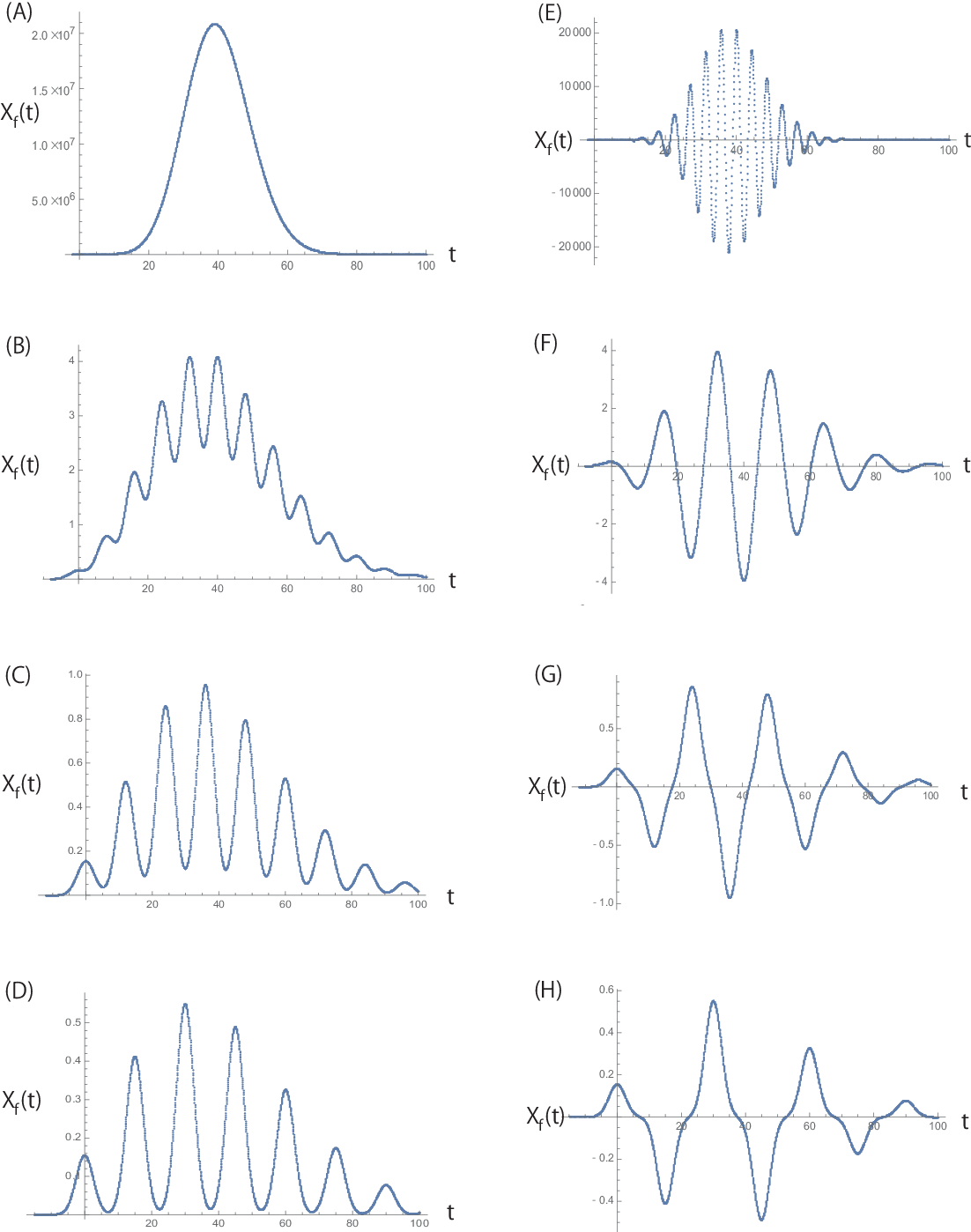}
\caption{Representative examples of solutions ${X_f}(t)$ given in (\ref{solx2}) through the Fourier transform. The value of ${\cal{C}}$ is set to $1$. Other parameters $(a, b, \tau)$ are configured as follows: (A)$(0.15, 6.0, 2.0)$, (B)$(0.15, 6.0, 8.0)$, (C)$(0.15, 6.0, 12.0)$, (D)$(0.15, 6.0, 15.0)$, (E) $(0.15, -6.0, 2.0)$, (F) $(0.15, -6.0, 8.0)$, (G) $(0.15, -6.0, 12.0)$, (H) $(0.15, -6.0, 15.0)$. }
\end{center}
\label{dynamics1}
\end{figure}

\section{Comparison with Numerical Solutions}

Returning to equation (\ref{ddr}) with $a>0$, we seek a real solution in the range $t \in [0, \infty)$ given the initial function $\phi (t)$, $t \in [-\tau, 0]$. Equation (\ref{ddr})  with a general initial function remains unsolvable. However, the earlier findings suggest the following conjecture:
\vspace{1em}

If the initial function 
$\phi(t)$ is represented by the solution 
${X_f}(t)$ derived in equation (\ref{solx2}) for 
$t \in [-\tau, 0]$, then the solution 
${X_s}(t)$ of equation (\ref{ddr}) is expressed by 
${X_f}(t)$ for 
$t \in [0, \infty)$. 
\vspace{1em}

In other words, if a specific initial function $\phi(t)$  is carefully selected such that it is the same as 
${X_f}(t)$ for $t \in [-\tau, 0]$, then the solution ${X_s}(t)$ for $t \in [0, \infty)$ is accurately given by ${X_f}(t)$ for $t \in [0, \infty)$.

To verify the effectiveness of the proposed method for constructing a specific solution through the Fourier transform, we conducted direct numerical integration of equation (\ref{ddr}). The initial condition was approximated by a 10th-order polynomial function: $\phi(t) \approx {\sum_{i=0}^{i=10}} d_i x^i $, chosen to closely match ${X_f}(t)$ for $t \in [-\tau, 0]$.
The corresponding results, aligned with those in Fig. 1, are illustrated in Fig. 2. 
For comparison, we also show Fig. 3, where the plots of Figs. 1 and 2 are overlaid.
\begin{figure}[h]
\begin{center}
\includegraphics[height=9.5cm]{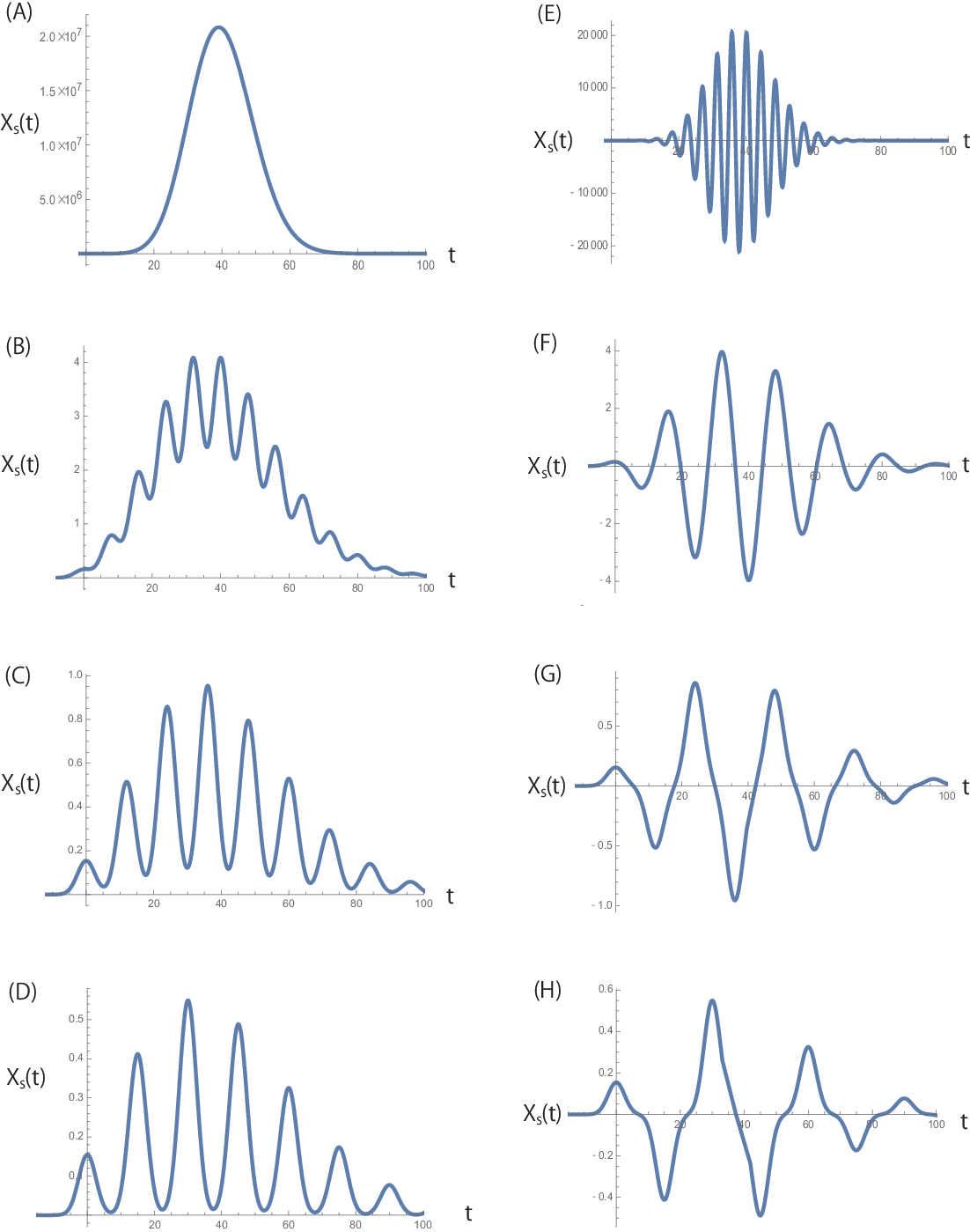}
\caption{Representative examples of solutions ${X_s}(t)$  obtained by the direct numerical integration of equation (1) with the initial function matching approximation. The value of ${\cal{C}}$ is set to $1$. Other parameters are identical to those in Fig. 1. The configurations for $(a, b, \tau)$ are as follows: (A)$(0.15, 6.0, 2.0)$, (B)$(0.15, 6.0, 8.0)$, (C)$(0.15, 6.0, 12.0)$, (D)$(0.15, 6.0, 15.0)$, (E) $(0.15, -6.0, 2.0)$, (F) $(0.15, -6.0, 8.0)$, (G) $(0.15, -6.0, 12.0)$, (H) $(0.15, -6.0, 15.0)$.} 
\end{center}
\label{dynamics2}
\end{figure}
\begin{figure}
\begin{center}
\includegraphics[height=9.5cm]{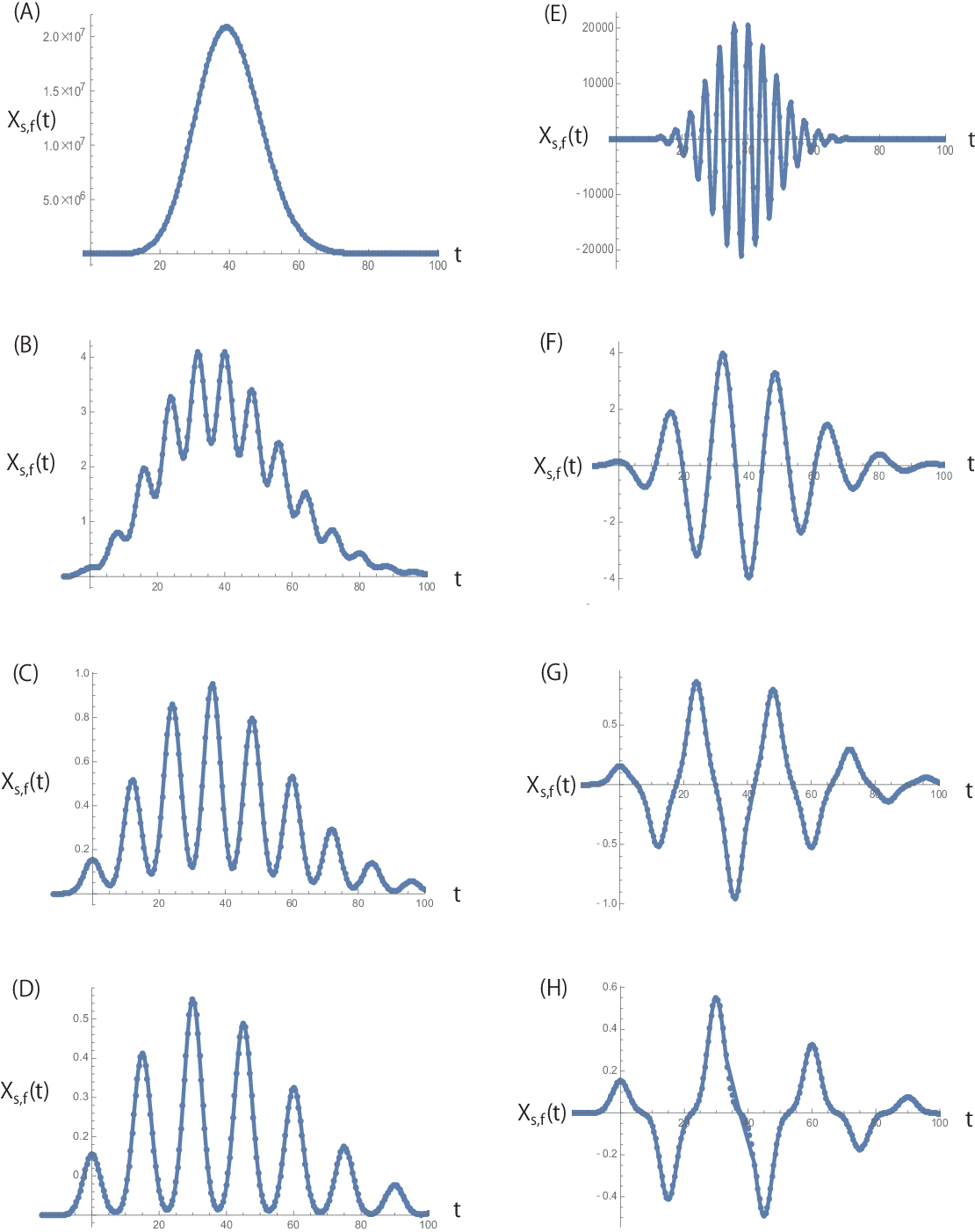}
\caption{Overlay plots of Fig. 1 and Fig. 2. Solid lines represent ${X_s}(t)$ from Fig. 2, and dots denote sampled points from ${X_f}(t)$ in Fig. 1. }
\end{center}
\label{dynamics3}
\end{figure}

Our observation indicates that the semi-numerical construction of the solution for equation (\ref{ddr}) through the Fourier transform can effectively capture the transient dynamics. This lends support to the reliability of our proposed approach.

\section{Discussion}

Several points warrant discussion regarding the findings presented.

1. Uncommon Capture of Transient Dynamics:
It is noteworthy that capturing transient or non-equilibrium dynamical trajectories is often challenging, not only in the context of delay differential equations but also in general nonlinear dynamical systems(e.g. \cite{cantisan,omel2022}. Despite being semi-numerical, having concrete ``solvable'' examples, as demonstrated in this study, can  contribute to a deeper understanding of non-linear transient phenomena.

2. Analytical Techniques and Challenges:
In the analysis of delay differential equations, the predominant tool is often the use of the Laplace transform or equivalent techniques. These approaches are valuable for revealing stability characteristics of solutions. However, in the case of equation (\ref{ddr}), the use of the Laplace transform does not yield an explicit solution, as achieved through the Fourier transform. The potential of further developing the Fourier transform approach to gain insight into the transient dynamics of general delay differential equations remains an open area for exploration.

3. Quantum Mechanical Connections:
Another intriguing observation is that equation (\ref{ddr}) with 
$b=0$ represents the annihilation operator equation for Quantum harmonic oscillators. Further differentiation of (\ref{ddr}) with respect to time yields a second-order delay differential equation. Exploring this direction and potential connections to Quantum mechanics is left for future investigations.

In conclusion, the findings presented in this study offer a potential pathway to advancing our comprehension of delay differential equations. The application of Fourier transform techniques demonstrates promise in capturing transient non-linear dynamics. Also, the identified connections to Quantum mechanics present an intriguing direction for future research and exploration.
\vspace{1em}

\noindent
{\bf Acknowledgments}

The authors would like to thank Prof. Hideki Ohira and the members of his research group at Nagoya University for their useful discussions. This work was supported by the ``Yocho-gaku'' Project sponsored by Toyota Motor Corporation, JSPS Topic-Setting Program to Advance Cutting-Edge Humanities and Social Sciences Research Grant Number JPJS00122674991, JSPS KAKENHI Grant Number 19H01201, and the Research Institute for Mathematical Sciences, an International Joint Usage/Research Center located at Kyoto University.

\end{document}